\documentclass[11pt]{article}
\usepackage{graphics}
\voffset - 1.0in
\title{PI-PI SCATTERING LENGTHS IN THE LIGHT OF PRECISION MEASUREMENTS}
\author{JACK L. URETSKY \\ High Energy Physics Division,
Argonne National Laboratories}
\date{\today }
\begin{document}\setlength{\unitlength}{.7mm}
\hfill ANL-HEP-PR-06-62
\today
\maketitle
\section{ABSTRACT}
I summarize the history of theoretical predictions of, and experimental attempts to measure, pion-pion (S-wave) scattering lengths.  Recent measurements at CERN confirm Weinberg's 1966 prediction of the I=2 scattering length and Basdevant' and Lee's subsequent correction of the Weinberg I=0 value by inclusion of an S-wave I=0 resonance. 
\section{INTRODUCTION}
	The Dirac Collaboration at CERN has recently announced its most recent results for the lifetime of the pionium atom, equivalent to a measurement of the pion-pion scattering lengths (the subscripts denote I-spin):\\      
                                 $|a_{2}-a_{0}| =0.280 \pm 0.015 \:  m_{\pi}^{-1}$ \cite{B. Adeva 07}\\
The Dirac result agrees with the measurement by the NA48/2 collaboration \cite {Batley} based on observation of a cusp-like structure in the $\pi ^{0} -\pi ^{0}$mass distribution seen in $~2.3\times 10^{7}$ $K^{\pm } \rightarrow \pi ^{\pm } \pi ^{0}\pi^{0}$ decays
                                 $|a_{0}-a_{2}| =0.268 \pm  0.010(sta) \pm 0.004(syst) \pm 0.013(external) \:  m_{\pi}^{-1}$  and $a_{2} = -0.041 \pm 0.022(sta)\pm 0.014(syst)\:  m_{\pi}^{-1}$
The ``external'' uncertainty refers to  uncertainty in the weak decay amplitudes,
         
         A recent prediction of the two scattering lengths, based upon chiral perturbation theory, \cite{G. Colangelo}  is   $a_{0}= 0.220 \pm    0.005 \; m_{\pi}^{-1} \:  \, a_{2}=-0.0444  \pm  0010  \;m_{\pi}^{-1}$
         ,   The Dirac Collaboration has submitted proposals to improve their measurement to an accuracy of 1 per cent in the scattering length difference . \cite{Schweizer}

           There are a number of lattice gauge calculations of $a_{2}$ reported in the literature.  Some of these are reviewed by the NPLQCD Collaboration \cite{NPLQCD} which reports their own mixed-action lattice result as $-0.04330\pm .00042\:m_{\pi}^{-1}$.  Not included in their summary is  a report of a calculation with improved actions on an anisotropic lattices by a Peking University collaboration \cite{Peking} giving a value of $-0.0467(45) \: m_{\pi}^{-1}$.
           
           These results may be compared with the soft-pion results obtained by Weinberg in 1966,
 namely $a_{0}=0.1588\:  m_{\pi}^{-1} \:\:a_{2}=-0.04537\:  m_{\pi}^{-1}$ (as calculated in \cite{NPLQCD}).
 
	This history, which is displayed graphically in the accompanying Figure 1,  makes evident that  low energy I=2 scattering was pretty well understood, at least from the time of Weinberg's calculation \cite{Weinberg}, but that the low energy I=0 scattering seemed anomalously large compared with current-algebra estimates.  It  seems, in fact, to have been a major motivation for chiral perturbation theorists, see, {\em e.g.}, ref \cite{G. Colangelo} to try to explain this difference.
	A purely hadronic explanation of the difference between the experimental and soft pion values of $a_{0}$ was given by Basdevant and Lee in 1970. \cite{B&L}   Their explanation may be understood by recalling that the scattering length for an attractive potential becomes large when there is a virtual bound state, that is, a resonance, at a low energy above threshold.  The Weinberg soft-pion result can be roughly simulated by a potential due to the exchange of a $\rho -$like particle (a particle that deviates somewhat from the mass and width of the actual $\rho $).  The difference between the soft-pion prediction and actual I=0 scattering length is then accounted for by the presence of a resonance in the I=0 S-wave channel.  This resonance, in the model of reference \cite{B&L} is the $\sigma$  (now known as the $f_{0}(600)$ of the $\sigma -$model of that reference.  The resonance is broad, and corresponds to a pole in the scattering amplitude at about 425 MeV.  The partial-wave amplitude was unitarized by use of Pade' -approximants.  The calculation also yielded resonances in the I=1, J=1 and I=0, J=2 channels, which the authors identified with the $rho $ and $f_{2}(1270)$.
	The pole position of the '``sigma'' found by Basdevant and Lee is within $1\frac{1}{2}$ standard deviations of the range of pole positions recently proposed by Colangelo, {\em et al.}\cite{G. Colangelo} based upon their chiral perturbation theory calculations.(see, especially, the discussion on p. 36 of the archived version of that reference).
	
\section{DISCUSSION}
	The striking feature of this history is  the robustness of the original Weinberg (``soft pion'') scattering length predictions \cite{Weinberg}, faced with the challenge of the increasingly precise experimental results.   From a strictly hadronic point of view it is as though inter-pionic forces are correctly given to a good approximation in the tree approximation by a simple effective Lagrangian \cite{Weinberg2}.  The I=0 S-wave calculated from this Lagrangian must be augmented, as in \cite{B&L}, by the inclusion of coupling to the $\sigma $ and, of course, to other resonances that couple to pion pairs in other partial waves.
	
	There have been attempts, not always successful, to calculate some of these resonant states as consequences of hadronic dynamics, see {\em e.g.} \cite{S&U}.  Now, in hindsight, it seems evident that these states must exist as a consequence of quark-gluon dynamics, and, in due course, they must emerge from the lattice.
	
	This history,  perhaps,  suggests an analogy between QED and QCD; just as we can do most, but not all, chemistry without knowing about QED, perhaps we can do most, but not all, hadronics without knowing about QCD.  We can expect, however, that the location of the boundary between hadronics and QCD will be further clarified by forthcoming high precision measurements of pi-pi scattering lengths in future Dirac Collaboration and similar measurements. 

\section{ACKNOWLEDGMENTS} 
I am indebted to Cosmas Zachos for several helpful discussions.   Work in the High Energy Physics Division at Argonne is supported by The U. S. Department of Energy, Division of High Energy Physics, Contract No. W-31-109-ENG-38
{\bf Note added after posting to the arXiv}
I had overlooked the results from new E865 Ke4 experiment at Brookhaven \cite{E865}. The experiment has subsequently been reanalyzed \cite{Descotes}, see also \cite{Gcolan2}. In my view, the process of deducing scattering lengths from final state interactions, as is done with the E865 data, is less credible than the more direct kind of measurements reported this year.  I am indebted to Professor Descotes for bringing the E865 data to  my attention and for some interesting comments.  I am also indebted to Professor Gerrold Franklin for reminding me of his analysis of an earlier Ke4 experiment \cite{Franklin}.

\begin{figure}
\includegraphics{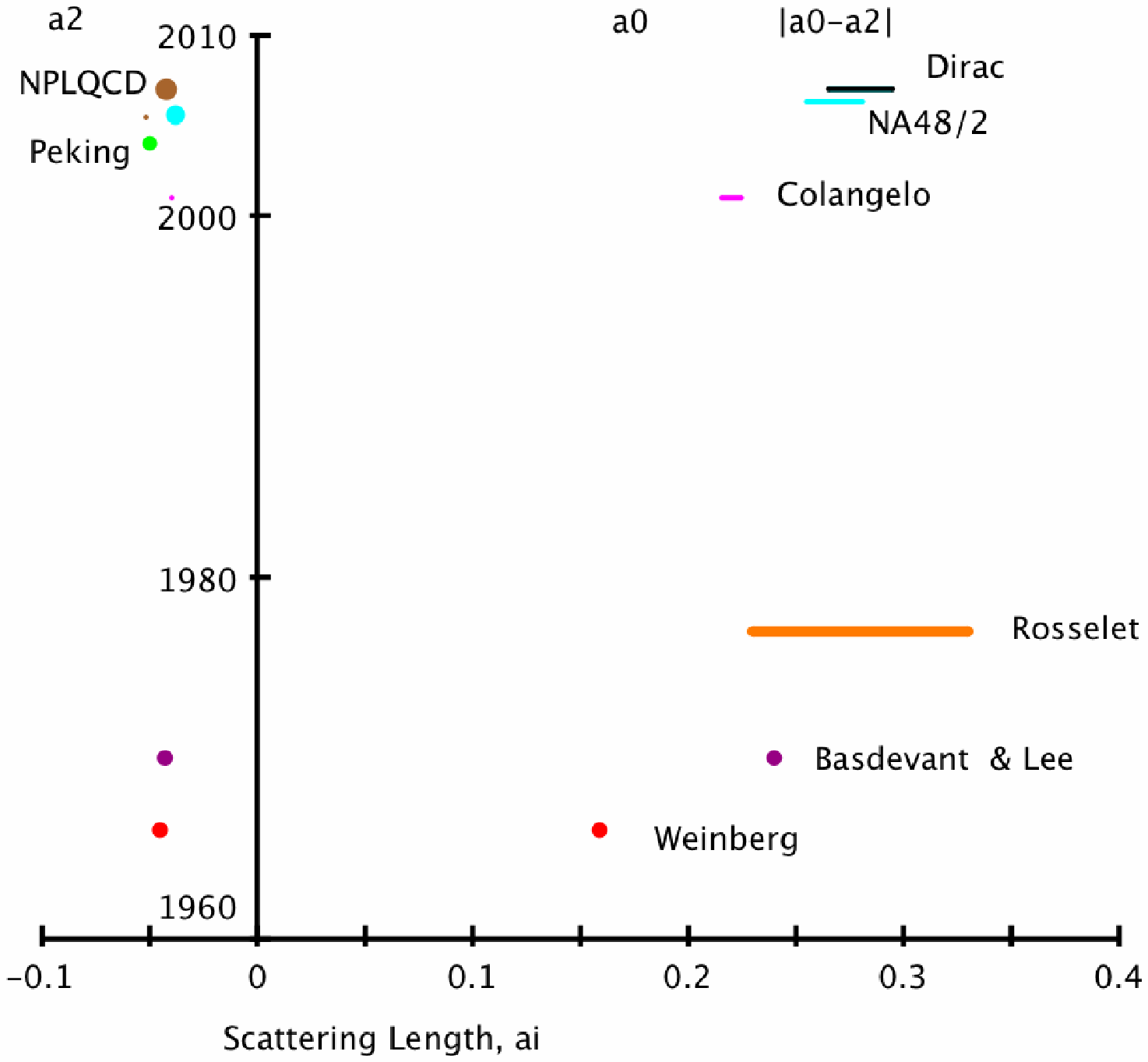}
 \caption{pi-pi scattering lengths, theory and experiment since 1960;  references are to the bibliography}
 \end{figure}
 \end{document}